\documentclass[aps,byrevtex,showpacs,showkeys,floats,floatfix,preprint]{revtex4}

\usepackage{amsmath,amssymb}
\usepackage{graphicx}
\usepackage{dcolumn}
\usepackage{bm}
\usepackage{srctex} 

\newcommand{\sig}[1]{\bm{\sigma}_{#1}}  
\newcommand{\Isig}[1]{I\!\bm{\sigma}_{#1}} 
\def\eps#1{\varepsilon_{#1}} 

\def\D{\cdot} 
\def\w{\!\wedge\!} 
\def\sig#1{\bm{\sigma}_{#1}} 
\def\Isig#1{I\!\bm{\sigma}_{#1}} 
\def\g#1{\gamma_{#1}} 

\def\bE{\mathbf{E}}
\def\bH{\mathbf{H}}
\def\bD{\mathbf{D}}
\def\bB{\mathbf{B}}
\def\bv{\mathbf{v}}

\def\cF{\mathcal{F}}
\def\cD{\mathcal{D}}
\def\cE{\mathcal{E}}
\def\cG{\mathcal{G}}
\def\cB{\mathcal{B}}
\def\cH{\mathcal{H}}

\begin{document}

\preprint{}

\title{Constitutive relations  for electromagnetic field in a form of $6\times 6$ matrices
derived from the geometric algebra}

\author{A.~Dargys}%
 \email{adolfas.dargys@ftmc.lt}
\affiliation{Center for Physical Sciences and Technology,
Semiconductor Physics Institute, Saul{\.e}tekio av.~3, LT-10222
Vilnius, Lithuania.}

\date{\today}

\begin{abstract}
To have a closed system,  the Maxwell equations should be
supplemented by constitutive relations which connect the primary
electromagnetic fields $(\bE,\bB)$  with the secondary ones
$(\bD,\bH)$ induced in a medium. Recently [Opt. Commun.
\textbf{354}, 259 (2015)] the allowed shapes of the constitutive
relations that follow from the relativistic Maxwell equations
formulated in terms of geometric algebra were constructed by
author. In this paper the obtained general relativistic relations
between $(\bD,\bH)$ and $(\bE,\bB)$ fields are transformed to four
$6\times 6$ matrices that are universal in constructing various
combinations of constitutive relations in terms of more popular
Gibbs-Heaviside vectorial calculus frequently used to investigate
the electromagnetic wave propagation in anisotropic, birefringent,
bianisotropic, chiral etc media.
\end{abstract}


\maketitle

\section{Introduction}\label{intro}

The Maxwell equations are not closed. The so-called constitutive
relations (CR) are needed to describe the electromagnetic (EM)
field propagation in a medium. The CR connects the pair
$(\bE,\bB)$ of electric and magnetic fields with the induced ones
$(\bD,\bH)$ in the medium where the wave propagates, for example
in a solid or plasma. A concrete form of the CR depends on
mathematics used to describe the fields. In Gibbs-Heaviside (GH)
analysis~\cite{Gibbs01,Arfken95}, where the fields are 3D vectors,
the constitutive relations between the fields are expressed by
$3\times 3$ matrices, for example, $\bD=\hat{\varepsilon}\bE$, or
in components $D_i=\sum_{j=1}^3 \varepsilon_{ij}E_j$, where
$\hat{\varepsilon}$ is the symmetric permittivity matrix with
$\varepsilon_{ij}=\varepsilon_{ji}$~\cite{Born99}. If relativistic
effects are important then one introduces the metric tensor and
goes either to tensor calculus~\cite{Post62}, which is popular
among physicists, or to dyadic  or GH calculus~\cite{Mackay10}
which are more popular within the engineering community.

A modern and more fundamental approach to electrodynamics is based
on differential p-forms~\cite{Hehl03} and Clifford geometric
algebras~\cite{Jancewicz88,Baylis99,Doran03,Arthur11,Dargys15}.
The former allows to formulate the premetric electrodynamics and
perform initial calculations without defining the space-time
metric at all. The GA requires to specify the metric to be used
but, due to 8-periodicity GA theorem and natural inclusion of
lower dimensional algebras and spaces  as one goes to higher
dimensional spaces, even for spaces characterized by different
metrics, the GA automatically  connects the classical and
relativistic calculations in a simple manner. For electrodynamics
the two of GA's are the most important, namely,
\textit{Cl}$_{3,0}$ which is related with the Euclidean 3D space
and \textit{Cl}$_{1,3}$ (sometimes \textit{Cl}$_{3,1}$) which
describes the Minkowski 4D space-time and contains
\textit{Cl}$_{3,0}$ as a subalgebra. In the former,
\textit{Cl}$_{3,0}$ algebra, the fields $\bE$ and $\bD$ are
represented as vectors while $\bB$ and $\bH$ are represented as
oriented planes, i.~e. planes that are spanned by two vectors
$\mathbf{a}$ and $\mathbf{b}$ and have two orientations which are
encoded in the outer product, $\mathbf{a}\w\mathbf{b}$ and
$\mathbf{b}\w\mathbf{a}=-\mathbf{a}\w\mathbf{b}$. In accordance
with \textit{Cl}$_{3,0}$, the electric field vector and magnetic
field bivector of a propagating EM wave in the vacuum lie in a
single plane, called the polarization plane. In contrast,
according to GH vector analysis~\cite{Gibbs01,Born99}, where the
magnetic field is treated as an axial vector, there arise
interpretational difficulties. Since in this interpretation the
electric and magnetic fields are perpendicular to EM wave
propagation direction (Poynting vector) there appears ambiguity as
to which of the fields (electric or magnetic) represents the true
polarization of EM wave in the vacuum. Usually we identify the
polarization vector with electric field if the EM wave falls onto
dielectric. However, if EM wave penetrates into magnetic material
the polarization should be ascribed to the magnetic field since
now the polarization is totally controlled by the magnetic
susceptibility of the medium. Thus it appears that the
polarization of the EM wave in vacuum depends on material
properties onto which if falls.  The paradox arises due to
mathematical inconsistency, since the axial  vector is not a
member of the vectorial space where the Maxwell equations are
usually constructed.  In GA interpretation the magnetic field is
an oriented plane (bivector) rather than the vector. The different
behavior of $\bE$ and $\bB$ fields under reflection -- so
difficult to explain to the students -- is very simple to
understand if GA is appealed to. Since in 3D Euclidean space there
are three mutually perpendicular planes (which may be identified
with the bivectors if the arrows pointing  anticlockwise are drawn
on the planes), one may find ono-to-one correspondence between GH
axial vectors and \textit{Cl}$_{3,0}$ algebra bivectors. For this
purpose it is enough the  GH axial vector (magnetic field) to
replace by oriented basis plane which is perpendicular to the
axial vector.  This explains why in classical \textit{Cl}$_{3,0}$
interpretation the electric and magnetic fields lie in a single
bivector plane which may identified with the polarization. In
addition, the Poynting vector lies in the same oriented plane.
However, in 4D space one has four basis vectors and six basis
planes. As a result the concept of axial vector in relativity
theory  fails and one must start from the beginning with new
mathematics. In GA the polarization of EM wave is an integral part
of the respective algebra, thus no interpretational ambiguities
with the polarization arise. From GA point of view, the commonly
used GH vectorial calculus~\cite{Gibbs01,Arfken95} is nothing else
but a crippled quaternionic algebra. The quaternions belong to
\textit{Cl}$_{0,2}$ algebra, which in its turn is the subalgebra
of \textit{Cl}$_{3,0}$, while the latter in its turn is the
subalgebra of the relativistic \textit{Cl}$_{1,3}$ algebra. This
hierarchy and coherent mathematical notation allows to formulate
all physics, including the mechanics, electrodynamics, quantum
mechanics and gravitation theory, in a single mathematical
picture. However the most of physicists are still unaware of this
new kind of mathematics.

General forms of CR's formulated in terms of classical
\textit{Cl}$_{3,0}$ and relativistic \textit{Cl}$_{1,3}$  algebras
were deduced  in papers~\cite{Dargys15a,Dargys15b}. Since the GA
is not widely known to physicists the aim of this paper is to
display the constitutive relations that follow from space-time
structure encoded in the relativistic GA in a form of $6\times 6$
matrices which can be easily transformed to Gibbs-Heaviside form
and applied to investigate properties of EM waves in various
media. Before  presenting the matrices it may be useful to list
some of essential properties of \textit{Cl}$_{1,3}$ algebra that
will help the reader to grasp how the relativity and CR's are
built in this GA.

1) \textit{Cl}$_{1,3}$ is constructed from four orthogonal vectors
$\g{i}$ which define the basis vectors in the space-time. $\g{0}$
is the time axis and $\g{1}$, $\g{2}$, and $\g{3}$  are the space
axes. The algebra of vectors $\g{i}$ are isomorphic to algebra of
Dirac matrices usually used in the relativistic quantum mechanics.
The squares of $\g{i}$'s satisfy $\g{0}^2=1$ and
$\g{1}^2=\g{2}^2=\g{2}^2=-1$ and thus define  $(+,-,-,-)$  metric
of the space-time.

2) Apart from the vectors the space-time contains  more geometric
objects:  bivectors $\g{i}\w\g{j}\equiv\g{ij}$ (six oriented
planes), trivectors $\g{i}\w\g{j}\w\g{l}$ (four oriented 3D
volumes) and the pseudoscalar which is equal to  outer products of
all basis vector, $I=\g{0}\w\g{1}\w\g{2}\w\g{3}$.  The EM field in
GA is represented by a general bivector plane that can be
decomposed into six projections, or six basis bivectors
$\g{i}\w\g{j}$, $i\ne j$. The number of basis bivectors is equal
to the number of EM field components in relativistic
electrodynamics. Three bivectors $\sig{i}\equiv\g{i}\w\g{0}$,
where $\g{0}$ is the time coordinate, are time-like, i.~e. odd
with respect to spatial inversion, and are connected with the
electric field. Their squares are $\sig{i}^2=+1$. The remaining
bivectors $\Isig{1}=\g{3}\w\g{2}$, $\Isig{2}=\g{1}\w\g{3}$ and
$\Isig{3}=\g{2}\w\g{1}$ that represent the magnetic field are
space-like (even with respect to spatial inversion). Their squares
are negative, $\big(\Isig{i}\big)^2=-1$.

Thus, in \textit{Cl}$_{1,3}$ algebra the primary EM field
$\cF=\cE+\cB$  called the Faraday bivector can be decomposed into
six elementary bivectors (projections) that represent six oriented
planes in the 4D Minkowski space-time:
\begin{eqnarray}
\cE&=& E_1\sig{1}+E_2\sig{2}+E_3\sig{3},\qquad\cE^2 >0,\label{elfield} \\
\cB&=& B_1\Isig{1}+B_2\Isig{2}+B_3\Isig{3},\quad\cB^2
<0.\label{magfield}
\end{eqnarray}
The real-valued coefficients before time- and space-like basis
bivectors mathematically  can be obtained by relativistic
space-time splitting operation~\cite{Doran03}.  They represent the
projections of 3D electric and magnetic fields $\bE=(E_1,E_2,E_3)$
and $\bB=(B_1,B_2,B_3)$ that are accessible to experiment.

3) The important property of electrodynamics formulated in GA
terms  is that there is no need for additional space-time symmetry
considerations. The automorphisms or involution symmetries of GA,
namely, the identity, inversion, reversion and Clifford
conjugation  are isomorphic to discrete Gauss-Klein group of four
$\mathbb{Z}_2\otimes\mathbb{Z}_2$~\cite{Varlamov01}, which in its
turn is  isomorphic to  group consisting of identity operation,
space~$P$ and time~$T$ reversals, and the combination~$PT$. Thus
the discrete  space-time symmetry operations
$\mathbb{Z}_2\otimes\mathbb{Z}_2\cong\{1,P,T,PT\}$ are satisfied
in GA  automatically. The idea that involutions define physically
important subspaces was expressed for the first time  by J.~Dauns
in his article entitled ``Metrics are Clifford algebra
involutions"~\cite{Dauns88}.

From all this follows that the multiplicative CR's that connect
primary and secondary EM fields  are integrated in relativistic
\textit{Cl}$_{1,3}$ algebra and there is no need for additional
assumptions. However, the additive CR's such as spontaneous
electric $\mathbf{P}$ and magnetic $\mathbf{M}$ polarizations are
not included in the results presented below. The reader who is
unfamiliar with the GA may go directly to final results, i.e., to
matrices~\eqref{epsmu}, \eqref{Fizeau}, \eqref{Fara} and
\eqref{optact} which may be used to construct various CR's in a
more conventional GH form.

\section{General form of constitutive relations for EM fields\label{sec:2}}
We shall assume that the medium is lossless, linear, and
unbounded, with instantaneous response to external fields. Then
the constitutive relation between the Faraday field $\cF$ and
excitation field $\cG$ induced in the medium is
\begin{equation}\label{constRelat}
\cG=\cD+\cH=\chi(\cE+\cB)=\chi(\cF),
\end{equation}
where the operator~$\chi$ is a linear bivector-valued function of
the bivector argument.  Also, we shall assume that CR between
$\cF$ and $\cG$ is local. The Maxwell equations in
\textit{Cl}$_{1,3}$ written in terms of primary $\cF$ and
secondary $\cG$ fields can be found in~\cite{Doran03}. We shall
adopt that the vacuum constants are normalized,
$\varepsilon_0=\mu_0=1$, so that Eq.~\eqref{constRelat} is
dimensionless. Conversion to dimensional form was given
in~\cite{Dargys15b}.

If $\cF$ and $\cG$ are columns that represent  component of the
bivectors  in the order
$(\sig{1},\sig{2},\sig{3},\Isig{1},\Isig{2},\Isig{3})$ than the
relation between the primary and secondary fields
in~\eqref{constRelat} can be expressed through  $6\times 6$ matrix
$\hat{\chi}$ with elements
$\hat{\chi}_{ij,kl}=\g{ij}\D(\hat{\chi}\,\g{kl})$, where $ij$ and
$kl$ are compound indices that run from $1$ to $6$, and  where the
dot means the inner GA product. Then the
transformation~\eqref{constRelat}  becomes
\begin{equation}\begin{split}\label{transform}
 \left[\begin{array}{c}
\cD\\
\cH\\
\end{array}\right]&=\Big(
\left|\begin{array}{c|c}
\text{Diel.Birefr.}&\text{Fizeau}\\
\hline
\text{Fizeau}&\text{Magn.Birefr.}\\
\end{array}\right|
+\\
&\quad\quad\left|\begin{array}{c|c}
\text{Diel.Faraday}&\text{Opt.Act.}\\
\hline
\text{Opt.Act.}&\text{Magn.Faraday}\\
\end{array}\right|
\Big) \left[\begin{array}{c}
\cE\\
\cB\\
\end{array}\right].
\end{split}\end{equation}
$3\times 3$ submatrices in~\eqref{transform} are named according
to some characteristic physical effect they represent. The first
$6\times 6$  matrix is symmetric. Its diagonal blocks are related
with dielectric and magnetic birefringence. The off-diagonal
blocks mix electric and magnetic fields and are responsible for
the Fizeau effect. The second asymmetric  matrix is responsible
for dielectric and magnetic Faraday effects and optical activity.
The division into two, symmetric and antisymmetric matrices
in~\eqref{transform}, is related with the adjoint transformation
in GA (analogue of transpose in the matrix notation under which
matrix rows and columns are interchanged with an appropriate
sign). Since an arbitrary adjoint transformation in GA can be
written as a sum of  symmetric and antisymmetric
transformations~\cite{Doran03}, the two matrices
in~\eqref{transform} correspond to this partition. The
constitutive relation is an operator $\chi$ which acts on the
bivector  $\cF$  and returns new bivector $\cG=\chi(\cF$) in the
Minkowski space. In the second, antisymmetric matrix
of~\eqref{transform}, after transformation, in addition, the
orientation of the new bivector $\cG$ is changed to opposite. One
can also imagine that opposite surfaces of the resulting plane
$\cG$ have been interchanged. This is a geometric content of the
CR that follows from \textit{Cl}$_{1,3}$ algebra. The described
linear GA transformation can be cast into $6\times 6$ matrix that
describes how the primary EM field components go to secondary
field components.

In the following we shall rewrite the equation~\eqref{transform}
in terms of coefficients that correspond to transformations within
or between time-like $\sig{}$ and space-like  $\Isig{}$ triads of
elementary bivectors
\begin{equation}\begin{split}\label{transform2}
& \left[\begin{array}{c}
\sig{}\\
\Isig{}\\
\end{array}\right]=\Big(
\left|\begin{array}{c|c}
\eps{}&\gamma^{s},\gamma^{a}\\
\hline
\gamma^{s},\gamma^{a}&\mu^{-1}\\
\end{array}\right|_{\textrm{sym}}
+\\
&\quad\quad\left|\begin{array}{c|c}
n&s^s,s^a\\
\hline
s^s,s^a&m\\
\end{array}\right|_{\textrm{antisym}}
\Big) \left[\begin{array}{c}
\sig{}\\
\Isig{}\\
\end{array}\right],
\end{split}\end{equation}
where various effects were replaced by respective symbols they
represent. In short, the individual submatrices
in~\eqref{transform2} interconnect general time-like $\sig{}$ and
space-like $\Isig{}$ GA bivectors. If only the upper-left block of
the first ``sym'' matrix is taken into account then we have a
constitutive relation that connects time-like bivector with
time-like bivector, i.~e. we have a symmetrical
$(\sig{},\sig{})_s$ coupling between primary and secondary fields.
There are more couplings such as $(\sig{},I\sig{})_s$,
$(I\sig{},\sig{})_s$, and $(\Isig{},\Isig{})_s$  that will
represent different linear electromagnetic and optical  effects.
Similarly, the second, antisymmetric   part provides a set of
bivector couplings $(\sig{},\sig{})_a$, $(\sig{},I\sig{})_a$,
$(\Isig{},\sig{})_a$, and $(\Isig{},\Isig{})_a$ which represent
different physical effects and which have, as mentioned,  opposite
orientations of secondary EM field bivectors.

\section{Symmetric part}
The symmetric part consists of three different  $3\times 3$
submatrices that correspond to $(\sig{},\sig{})_s$,
$(\Isig{},\Isig{})_s$ and $(\sig{},\Isig{})_s=(\Isig{},\sig{})_s$
couplings.

\paragraph{$(\sig{},\sig{})_s$ and $(\Isig{},\Isig{})_s$ couplings
--- electrical and magnetic birefringence.}
The coupling $(\sig{},\sig{})_s$  describes the effect of
anisotropic dielectric on light propagation, for example the
birefringence of light in a quartz. The permittivity submatrix
$\hat{\varepsilon}$ generated by GA transformation is symmetric,
$\eps{ij}=\eps{ji}$, and consists of 6 independent scalars which
connect two time-like bivectors, $\cE$ and $\cD$, as the
transformation symbol $(\sig{},\sig{})_s$ implies.

Similarly, in the $(\Isig{},\Isig{})_s$ coupling  the space-like
bivector $\cB$ is transformed to other space-like bivector $\cH$.
It corresponds to the inverse permeability matrix $\hat{\mu}^{-1}$
which is characterized by  six scalars too. So the compound
transformation  describes the bianisotropic medium which
transforms the EM field $\cF=\cE+\cB$ to
$\cD+\cH=\chi_{\varepsilon,\mu^{-1}}(\cF)$. The  GA transformation
operator
$\chi_{\varepsilon,\mu^{-1}}=\chi_{\varepsilon}+\chi_{\mu^{-1}}$
yields the matrix $\hat{\chi}_{\varepsilon,\mu^{-1}}$ with
elements $\g{ij}\D\chi_{\varepsilon,\mu^{-1}}(\g{kl})$,
\begin{equation}\label{epsmu}
\hat{\chi}_{\varepsilon,\mu^{-1}}=\left[\begin{array}{cccccc}
\eps{11}&\eps{12}&\eps{13}&0&0&0\\
\eps{12}&\eps{22}&\eps{23}&0&0&0\\
\eps{13}&\eps{23}&\eps{33}&0&0&0\\
0&0&0&\mu_{11}^{-1}&\mu_{12}^{-1}&\mu_{13}^{-1}\\
0&0&0&\mu_{12}^{-1}&\mu_{22}^{-1}&\mu_{23}^{-1}\\
0&0&0&\mu_{13}^{-1}&\mu_{23}^{-1}&\mu_{33}^{-1}
\end{array}\right].
\end{equation}
It is understood that this matrix acts on the column-vector
$(E_1,E_2,E_3,B_1,B_2,B_3)^\text{T}$, or in short
$(\bE,\bB)^{\text{T}}$ where $\text{T}$ means ``Transpose''. The
result is the excitation column-vector $(\bD,\bH)^\text{T}$.

\paragraph{$(\sig{},\Isig{})_s$ coupling --- Fresnel-Fizeau
effect.}  This coupling transforms the space-like bivector to
time-like bivector, $\sig{i}\rightarrow\Isig{j}$, or vice versa,
$\Isig{i}\rightarrow\sig{j}$.  Since the structure of upper-right
and lower-left Fresnel-Fizeau submatrices in \eqref{transform2} is
similar (the submatrices  are antisymmetric with respect to main
diagonal), the structure of both transformations,
$\sig{}\to\Isig{}$ and $\Isig{}\to\sig{}$, is similar too.  In its
turn, the individual $3\times 3$ submatrices may be decomposed
into sum of even and odd parts, consequently the transformation
$\cG_{\gamma}=\chi_{\gamma}(\cF)$ can be divided into sum of
matrices
$\hat{\chi}_{\gamma}=\hat{\chi}_{\gamma}^s+\hat{\chi}_{\gamma}^a$,
where the superscripts  $s$ and $a$ indicate the symmetric and
antisymmetric parts. Thus we find that the most general
transformation which is allowed by GA can be rewritten in the
following matrix form~\cite{Dargys15b},
\begin{widetext}
\begin{equation}\label{Fizeau}
\Hat{\chi}_{\gamma}=\left[
\begin{array}{cccccc}
0&0&0&\g{11}^s&\g{12}^s+\g{12}^a&\g{13}^s-\g{13}^a\\
0&0&0&\g{12}^s-\g{12}^a&\g{22}^s&\g{23}^s+\g{23}^a\\
0&0&0&\g{13}^s+\g{13}^a&\g{23}^s-\g{23}^a&\g{33}^s\\
\g{11}^s&\g{12}^s-\g{12}^a&\g{13}^s+\g{13}^a&0&0&0\\
\g{12}^s+\g{12}^a&\g{22}^s&\g{23}^s-\g{23}^a&0&0&0\\
\g{13}^s-\g{13}^a&\g{23}^s+\g{23}^a&\g{33}^s&0&0&0\\
\end{array}
\right].
\end{equation}
\end{widetext}
The matrix~\eqref{Fizeau} represents various nonreciprocal
effects. All in all, it contains 9 independent scalar parameters,
three of which $(\g{12}^a,\g{13}^a,\g{23}^a)$ belong to the skew
symmetric part of submatrices and can be represented by a vector.

The combined action of EM birefringence and Fizeau effects is
given by  sum of  \eqref{epsmu} and \eqref{Fizeau}. In a simple
case when the permittivity and permeability are   scalars, and the
coupling $(\sig{},\Isig{})_s$  has vectorial form  the sum
simplifies to
\begin{equation}\label{constmatc}
\hat{\chi}_s=\left[\begin{array}{cccccc}
\eps{}&0&0&0&\g{12}^a&-\g{13}^a\\
0&\eps{}&0&-\g{12}^a&0&\g{23}^a\\
0&0&\eps{}&\g{13}^a&-\g{23}^a&0\\
0&-\g{12}^a&\g{13}^a&\mu^{-1}&0&0\\
\g{12}^a&0&-\g{23}^a&0&\mu^{-1}&0\\
-\g{13}^a&\g{23}^a&0&0&0&\mu^{-1}
\end{array}\right].
\end{equation}
In the standard GH notation with the axial magnetic field vector
$\bB$ introduced, the matrix~\eqref{constmatc} leads to the
following constitutive relation between the fields
\begin{eqnarray}
\bD&= \hat{\varepsilon}\,\bE-\bm{\gamma}^a\times\bB, \label{velD}\\
\bH&=\bm{\gamma}^a\times\bE+\hat{\mu}^{-1}\bB, \label{velH}
\end{eqnarray}
where the cross indicates standard vectorial product and
$\bm{\gamma}^a=(\g{23},\g{13},\g{12})$ is the coupling vector. The
same form of the constitutive relations was obtained earlier using
simpler \textit{Cl}$_{3,0}$ algebra~\cite{Dargys15a}, which
describes the classical electrodynamics in 3D Euclidean space and
where time is a parameter rather than the space-time vector. In
\eqref{velD} and \eqref{velH} the anisotropy of the medium is
characterized by coupling vector $\bm{\gamma}^a$, which may be,
for example, the velocity $\bv$ of a fluid as in the Fizeau
experiment, or the velocity of a moving dielectric slab. However
it should be noted that the matrix~\eqref{Fizeau} was generated by
\textit{Cl}$_{1,3}$ algebra that describes the Minkowski
space-time  and therefore this matrix represents the relativistic
constitutive relation. The matrix~\eqref{Fizeau} can be applied
even when the velocity of medium approaches to the light velocity.
Finally, we shall remind that the total relativistic matrix
$\hat{\chi}_{\gamma}$ is symmetric while the asymmetry appears in
individual $3\times 3$ blocks only. In conclusion, the symmetric
matrices \eqref{epsmu} and \eqref{Fizeau}  represent
transformation that corresponds to adjoint symmetric GA
transformation. In GA terms it can be written as a sum
$\cG=(\cD+\cH)=\chi_{\textrm{sym}}\cF=(\chi_{\eps{}}+\chi_{\mu^{-1}}+\chi_{\gamma})\cF$,
where linear GA transformation operators  have their respective
matrix analogues
$\hat{\chi}_{\textrm{sym}}=\hat{\chi}_{\eps{}}+\hat{\chi}_{\mu^{-1}}+\hat{\chi}_{\gamma}$
in  equations \eqref{epsmu} and \eqref{Fizeau}.

\section{Antisymmetric part}
The second, antisymmetric term in~\eqref{transform2}, in addition,
takes into account the change of bivector  direction  after the
transformation, figuratively speaking,  due to interchange of
opposite surface colors of the oriented bivector plane. This 4D
space-time property of \textit{Cl}$_{1,3}$ which is encoded in the
automorphisms  has no analogy in the classical
electrodynamics~\cite{Dargys15a}.

\paragraph{$(\sig{},\sig{})_a$ and $(\Isig{},\Isig{})_a$ couplings
--- electric and magnetic Faraday effects.}
In the electric  Faraday effect characterized by
$(\sig{},\sig{})_a$ coupling,  the electric field $\cE$ induces an
excitation or displacement bivector $\cD$ of the form
\begin{equation}\label{sigsiga}
\begin{split}
\cD=&\chi_n(\cE)=(E_3n_2-E_2n_3)\sig{1}+\\
&(E_1n_3-E_3n_1)\sig{2}+ (E_2n_1-E_1n_2)\sig{3},
\end{split}
\end{equation}
where  $n_1$, $n_2$, $n_3$ are  material parameters  which in the
standard notation are elements of  $3\times 3$  skew symmetric
matrix.

In the magnetic Faraday effect characterized by
$(\Isig{},\Isig{})_a$ coupling  the magnetic field bivector  $\cB$
is transformed to excitation bivector $\cH$. Similarly, this
transformation can be expressed by material bivector
$m=m_1\Isig{1}+m_2\Isig{2}+m_3\Isig{3}$, where $m_i$ are the
scalars that are magnitudes of the projections of space-like
bivector and define the strength and angular properties of the
physical effect. If transformed to the coordinate form, this
transformation is similar to~\eqref{sigsiga}
\begin{equation}\begin{split}
\cH=&\chi_m(\cB)=(B_2m_3-B_3m_2)\Isig{1}+\\
&(B_3m_1-B_1m_3)\Isig{2}+ (B_1m_2-B_2m_1)\Isig{3}.
\end{split}\end{equation}

Both the Faraday electric and  magnetic  transformations can be
rearranged as a single matrix~\cite{Dargys15b}
\begin{equation}\label{Fara}
\hat{\chi}_{n,m}=\left[\begin{array}{cccccc}
0&-n_3&n_2&0&0&0\\
n_3&0&-n_1&0&0&0\\
-n_2&n_1&0&0&0&0\\
0&0&0&0&m_3&-m_2\\
0&0&0&-m_3&0&m_1\\
0&0&0&m_2&-m_1&0
\end{array}\right]
\end{equation}
which may  be cast into the standard GH vectorial form as two
constitutive relations,
\begin{equation}
\bD=\mathbf{n}\times\bE,\quad \bH=-\mathbf{m}\times\bB
\end{equation}
where $\mathbf{n}=(n_1,n_2,n_3)$ and $\mathbf{m}=(m_1,m_2,m_3)$
are the material vectors. \vspace{3mm}

\paragraph{$(\sig{},\Isig{})_a$ coupling --- optical activity.} In
the optical activity,  the electric field $\cE$ is converted to
magnetic excitation $\cH$. Similar conversion occurs between $\cB$
and  $\cD$.  Expression~\eqref{transform2} shows that the optical
activity is described  by two off-diagonal submatrices. As follows
from GA, both  submatrices should be equal but  have opposite
signs so that the optical activity in fact is characterized by a
single $3\times 3$ submatrix with 9 parameters which can be
expressed as a sum of symmetric and skew symmetric parts
$\hat{\chi}_s^s$ and $\hat{\chi}_s^a$. In a matrix form the
combined transformation
$\hat{\chi}_s=\hat{\chi}_s^s+\hat{\chi}_s^a$ has the following
structure,
\begin{widetext}
\begin{equation}\label{optact}
\hat{\chi}_s=\left[
\begin{array}{llllll}
0&0&0&s_{11}^s&s_{12}^s-s_{3}^a&s_{13}^s+s_{2}^a\\
0&0&0&s_{21}^s+s_{3}^a&s_{22}^s&s_{23}^s-s_{1}^a\\
0&0&0&s_{31}^s-s_{2}^a&s_{32}^s+s_{1}^a&s_{33}^s\\
-s_{11}^s&-s_{21}^s-s_{3}^a&-s_{31}^s+s_{2}^a&0&0&0\\
-s_{12}^s+s_{3}^a&-s_{22}^s&-s_{32}^s-s_{1}^a&0&0&0\\
-s_{13}^s-s_{2}^a&-s_{23}^s+s_{1}^a&-s_{33}^s&0&0&0\\
\end{array}
\right],
\end{equation}
\end{widetext}
where without losing notational generality it may be assumed that
$s_{ij}^s=s_{ji}^s$.

Then,  in the symmetric part $\hat{\chi}_s^s$ we can select 6
independent scalar components which geometrically can be
represented by material ellipsoid. In the skew symmetric part
$\hat{\chi}_s^a$ we can select 3 independent components and
represent them  by vector $\mathbf{s}^a=(s_1^a,s_2^a,s_3^a)$.
Thus, in the standard vectorial notation the transformation
matrix~\eqref{optact} can be represented as
\begin{eqnarray}
\bD&=\hat{\chi}_s^s\bB+\mathbf{s}^a\times\bB ,\label{optactA}\\
\bH&=\mathbf{s}^a\times\bE -\hat{\chi}_s^s\bE.\label{optactB}
\end{eqnarray}
The symmetric part $\hat{\chi}_s^s$ is related with the chirality
of  medium. If $s_{ij}^s=0$ when $i\ne j$ and
$s_{11}^s=s_{22}^s=s_{33}^s$ then we have a single chiral coupling
constant  between primary and secondary fields, the case which is
frequently met in the computer modelling of chiral materials. The
cross-product terms in \eqref{optactA} and \eqref{optactB} also
appear in the Fizeau effect, equations \eqref{velD} and
\eqref{velH}. However, in contrast to equations \eqref{velD} and
\eqref{velH},  now  we see that the coupling vector $\mathbf{s}^a$
in  equations \eqref{optactA} and \eqref{optactB} has the same
sign. This difference in signs brings about principally different
physical effects.

\paragraph{Axionic part.} Axionic  contribution~\cite{Hehl03} comes from the trace
of the full transformation matrix \eqref{transform2}, actually
from the trace of the symmetric part of $6\times 6$ matrix. The
respective CR is described by product of unit matrix and scalar
constant (trace)~$\alpha$. In GA where the metric is predetermined
the axionic term does not appear. It should be remarked that at
present it remains  unclear whether the axionic field exists at
all. The experimental attempts to detect this field and respective
particle (axion)  up till now were unsuccessful.~\cite{axion}

To sum up, the Clifford geometric algebra provides a coherent
picture of how the constitutive relations for a homogeneous EM
media originate from all possible  linear transformations between
time-like and space-like bivectors of relativistic
\textit{Cl}$_{1,3}$ algebra and automorphisms (involutions) of
this algebra without any need for additional assumptions on
space-time properties. In the paper~\cite{Dargys15b} the obtained
set of 36 independent real coefficients generated by
\textit{Cl}$_{1,3}$ algebra coincides with that found by E.~J.
Post~\cite{Post62} from space-time symmetry consideration using
the tensorial calculus. All possible transformations that follow
from internal GA structure here were cast into a form of $6\times
6$ matrices \eqref{epsmu}, \eqref{Fizeau}, \eqref{Fara} and
\eqref{optact} which in turn were separated into $3\times 3$
susceptibility submatrices according to physical effects they
describe.  The obtained  constitutive $3\times 3$ matrices were
found to be either symmetric (represented by ellipsoids) or skew
symmetric (represented by vectors). They  may be applied to
investigate EM wave propagation in optics and electrodynamics.

The obtained $6\times 6$ susceptibility matrices \eqref{epsmu},
\eqref{Fizeau}, \eqref{Fara} and \eqref{optact} are rather
general. Since they are related with the linear transformations
between primary $(\cE,\cB)$ and secondary $(\cD,\cH)$ fields, they
may be added in various combinations to produce a multitude of
physical effects in EM wave propagation. Since the geometric
algebra is real the coupling  coefficients may assume real values
only including zero, of course. After transformation  of
constitutive matrices to the GH notation the EM fields may be
treated as complex vectors as it is usually done when the EM wave
propagation equations are written directly in this notation.
However, the summation of smaller $3\times 3$ susceptibility
submatrices which, as mentioned, appear in the frequently  used GH
notation, must be in accord with the symmetry and summation rules
for larger $6\times 6$ matrices \eqref{epsmu}, \eqref{Fizeau},
\eqref{Fara} and \eqref{optact}  to avoid forbidden combinations
between pairs of polar $(\bE,\bB)$ and axial $(\bD,\bH)$ fields.



\begin{thebibliography}{10}

\bibitem{Gibbs01}
J.W. Gibbs and E.B. Wilson,
\newblock {\em Vector Analysis},
\newblock Charles Scribner's Sons, New York, 1901.
\newblock (Digitized for Microsoft Corporation by the Internet Archive in 2007)

\bibitem{Arfken95}
G.B. Arfken,
\newblock {\em Mathematical Methods for Physicists},
\newblock Academic Press, San Diego, 1995.

\bibitem{Born99}
M.~Born and E.~Wolf,
\newblock {\em Principles of Optics},
\newblock Cambridge University Press, Cambridge, 1999,
\newblock 7-th (expanded) edition.

\bibitem{Post62}
E.J. Post,
\newblock {\em Formal Structure of Electromagnetics},
\newblock North Holland Publishing Company, Amsterdam, 1962.

\bibitem{Mackay10}
T.G. Mackay and A.Lakhtakia,
\newblock {\em Electromagnetic Anisotropy and Bianisotropy: A Field Guide},
\newblock World Scientific Publishing, Singapore, 2010.

\bibitem{Hehl03}
F.W. Hehl and Y.N. Obukhov,
\newblock {\em Foundations of Classical Electrodynamics: Charge, Flux and
  Metric},
\newblock Birkh{\"a}user, Boston, 2003.

\bibitem{Jancewicz88}
B.~Jancewicz,
\newblock {\em Multivectors and Clifford Algebra in Electrodynamics},
\newblock World Scientific, Singapore, 1988.

\bibitem{Baylis99}
W.~Baylis,
\newblock {\em Electrodynamics: A Modern Geometric Approach},
\newblock Birkh{\"a}user, Boston, 1999.

\bibitem{Doran03}
C.~Doran and A.~Lasenby,
\newblock {\em Geometric Algebra for Physicists},
\newblock Cambridge University Press, Cambridge, 2003.

\bibitem{Arthur11}
J.W. Arthur,
\newblock {\em Understanding Geometric Algebra for Electromagnetic Theory},
\newblock John Wiley \& Sons, Hoboken-New Jersey, 2011.

\bibitem{Dargys15}
A.~Dargys and A.~Acus
\newblock {\em Clifford Geometric Algebra and its Applications},
\newblock Petro ofsetas, Vilnius, 2015. (In Lithuanian)

\bibitem{Dargys15a}
A.~Dargys,
\newblock {\em Lith.  J.~Phys.} {\bf 55}, 92 (2015).

\bibitem{Dargys15b}
A.~Dargys,
\newblock {\em Opt. Commun.}  {\bf 354}, 259 (2015).

\bibitem{Varlamov01}
V.V. Varlamov,
\newblock {\em Int. J. Theor. Phys.} {\bf 40}, 769 (2001).

\bibitem{Dauns88}
J.~Dauns,
\newblock {\em Int. J. Theor. Phys.} {\bf 27}, 183 (1988).

\bibitem{axion}
Wikipedia,
\newblock ``Axion''.

\end{thebibliography}

\end{document}